\documentclass[letterpaper]{jpconf}
\usepackage{graphicx}
\newcommand {\pT}{$p_{T}$}

\newcommand {\ttbar}{$t\bar{t}$}

\newcommand {\bbb}{$bb\bar{b}$}
\newcommand {\btautau}{$b\tau\tau$}

\newcommand {\htautau}{$\phi \to \tau \tau$}

\newcommand {\et}{$E_{T}$}

\newcommand {\mett}{\displaystyle{\not} E_{T}}
\newcommand {\tanbeta}{$\tan \beta$}
\newcommand {\taunn}{$NN_{\tau}$}

\begin{document}

\title{Search for neutral Higgs bosons decaying to tau pairs in association with b-quarks at the D0 Detector}

\author{Kenneth Herner, for the D0 Collaboration}

\address{University of Michigan}

\begin{abstract}
We report results from a search for neutral Higgs bosons decaying to tau pairs produced in association with a $b$-quark in 1.2 fb$^{-1}$ of
data taken from June 2006 to August 2007 with the D0 detector at Fermi National Accelerator Laboratory. The final state includes
a muon, hadronically decaying tau and jet identified as coming from a $b$-quark. We set cross section times branching ratio limits
on production of such neutral Higgs bosons $\phi$ in the mass range from 90 GeV/$c^{2}$ to 160 GeV/$c^{2}$. Exclusion limits are set at the 95\%\
Confidence Level for several supersymmetric scenarios.

\end{abstract}

\section{Introduction}
In models with supersymmetry (SUSY), the Higgs sector is expanded relative to the
standard model (SM).  In the minimal supersymmetric standard model (MSSM), two Higgs doublet fields are needed instead of one, 
and the resulting Higgs
particle spectrum consists of two neutral scalars, a single neutral
pseudoscalar and two charged scalars.  The ratio of the vacuum expectation values of
the two doublets is denoted $\tan \beta$, and the coupling to weak isospin $-1/2$
members of the fermion doublets becomes proportional to $\tan \beta$.  In the mass range we consider, 90-160 GeV/$c^{2}$, 
two of the three neutral higgs bosons will be 
nearly degenerate in mass and this overlap results in a factor
of 2 gain in effective cross section.
In the MSSM the dominant Higgs decay for a pseudoscalar Higgs mass ($m_{A}$) below $\approx$500 GeV/$c^{2}$ is to a pair of $b$
quarks, with $\tau$ pair decays occurring with a branching ratio of roughly
10\%.  However, the $\phi b\rightarrow b\overline{b}b$
final state suffers from a large multijet background while the $\phi b\to $\btautau\
channel offers a much cleaner final state, giving the two channels similar sensitivities.  Additionally, while the $p\bar{p}\to $\htautau\ process 
has a higher cross section than 
the $p\bar{p}\to \phi b \to $\btautau\ process, \htautau\ suffers from irreducible $Z \to \tau\tau$ background, while the additional $b$-jet in \btautau\ largely alleviates this problem, giving \btautau\ additional sensitivity if $m_{\phi} \sim M_{Z}$.

The D0 detector ~\cite{run1det,run2det,run2muon} has a central-tracking system, hermetic liquid-argon calorimeter, and outer muon system.  We perform the search in 1.2 fb$^{-1}$ of data taken from June 2006 to August 2007.  We search for $\phi b$ in the \btautau\ final state, with one tau decaying to a muon and the other to hadrons.  This offers the best balance between a large branching ratio and robust object identification in the detector.

\section{Event Selection and Background Estimation}

\begin{figure}[htbp]
\centering
\includegraphics[scale= 0.35]{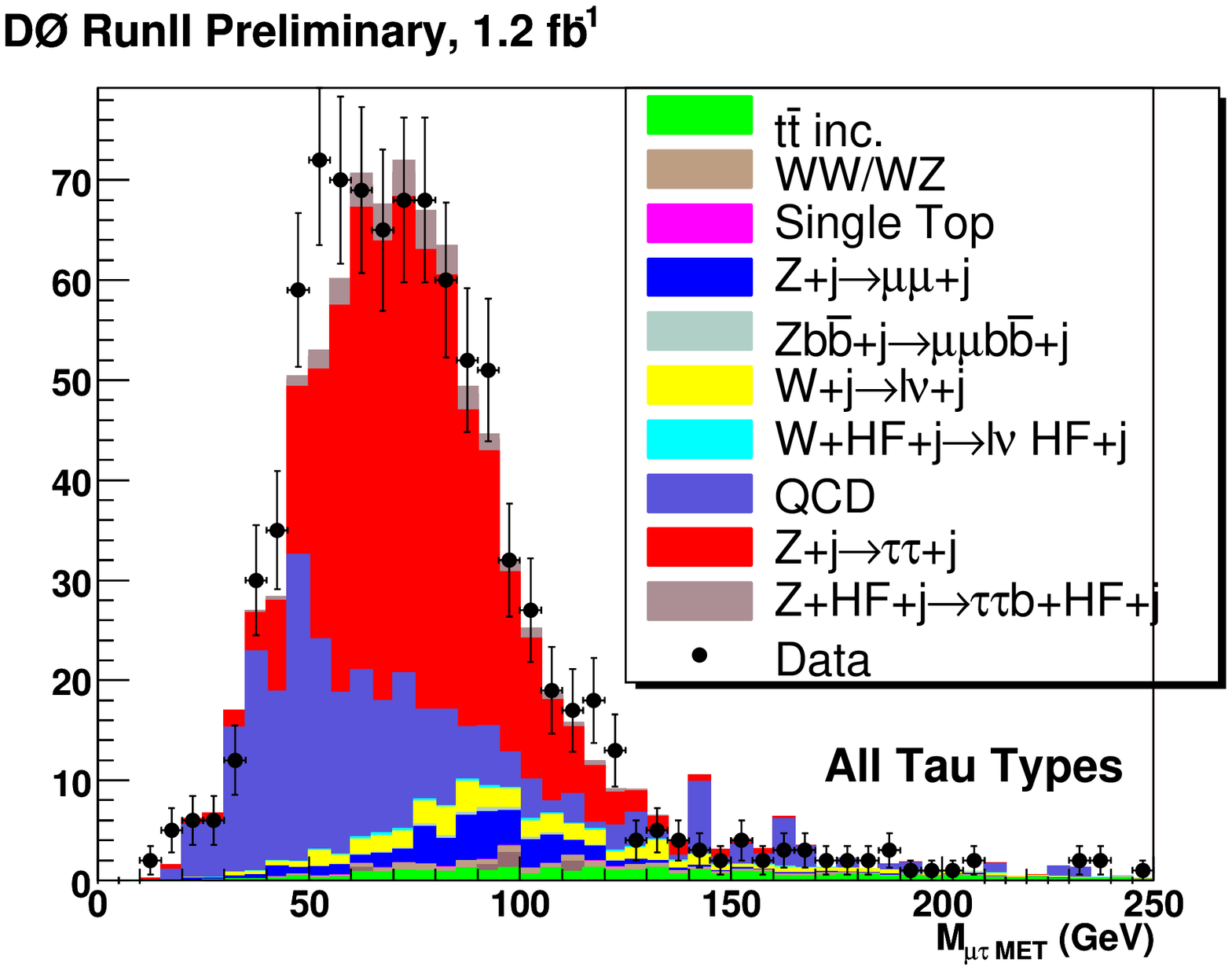}%
\includegraphics[scale= 0.35]{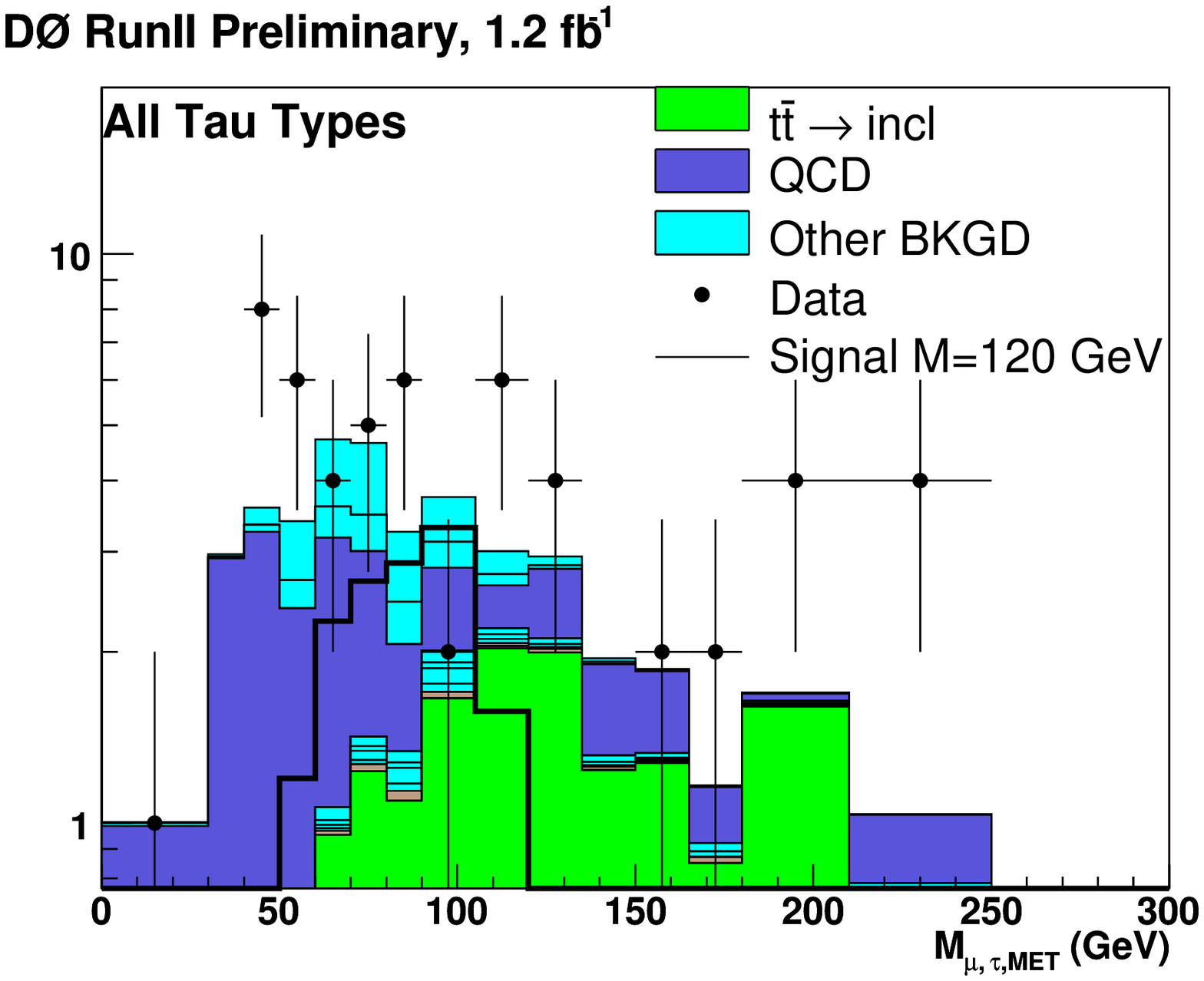}
\caption{Data/background comparison in the visible mass variable $(M(\mu,\tau,\mbox{MET})$ after muon, hadronic tau, and jet selection.  Left: before $b$-tagging.  Right: after $b$-tagging.  In the legend, ``QCD'' refers to the multijet background.\label{vismassfigure}} 
\end{figure}

The final state includes one muon, one hadronic tau candidate and one jet coming from $b$-quark fragmentation.  
require one isolated muon in the event with \pT\ $> 12$ GeV/$c$,
$\left|\eta\right| < 2.0$ and a central track match.  Hadronic taus appear as jet-like objects in the detector.  We apply a neural network (\taunn)\cite{taunn} to distinguish hadronic tau decays from jet fakes.  D0 has three different types of hadronic taus, differing in their number of charged tracks and/or the presence of electromagnetic energy in the calorimeter.  We make the following kinematic and neural network requirements on the hadronic tau candidate:

\begin{itemize}
\item Type 1: \et $ > 10$ GeV/$c$, $p_{T}^{trk} > 7$ GeV/$c$, \taunn $>0.9$
\item Type 2: \et $ > 10$ GeV/$c$, $p_{T}^{trk} > 5$ GeV/$c$, \taunn $>0.9$ 
\item Type 3: \et $ > 15$ GeV/$c$, 1 track with \pT$ > 5$ GeV/$c$, $\displaystyle
  \sum p_{T}^{trk} > 10$ GeV/$c$, \taunn $>0.95$.
\end{itemize}
Additionally we require at least one jet with \pT $> 15$ GeV/$c$, $|\eta|<2.5$.  At least one of the jets must be tagged as coming from $b$-quark fragmentation by the D0 neural network $b$-tagger.  Figure \ref{vismassfigure} shows the data/background comparison in the visible mass (defined as the invariant mass of the muon, hadronic tau, and missing transverse momentum) before and after $b$-tagging.

Backgrounds in this search include $W/Z$+jets production, multijet production, Di-boson, \ttbar  and single top production.  The $W/Z$+jets contribution is estimated using {\sc ALPGEN}\cite{b-alpgen} Monte Carlo (MC) interfaced with {\sc PYTHIA}\cite{pythia} for hadronization and showering.  Signal, Di-boson production and \ttbar\ production are estimated using {\sc PYTHIA}, and single top production via {\sc COMPHEP}\cite{comphep1,comphep2,comphep3}.  Multijet background is estimated primarily from data.  We use two independent methods to estimate multijet production; the first method relies on measuring the probability for a jet to be $b$-tagged in a multijet-enriched sample, while the second method uses the probability for a jet to pass the \taunn\ cut and for a muon to pass the isolation requirement in a multijet-enriched sample.  We take the average of the two methods as the final multijet contribution and include the difference between the two methods as a systematic error.  

\section{Multivariate Methods}

\begin{figure}[htb]
\includegraphics[scale=.28]{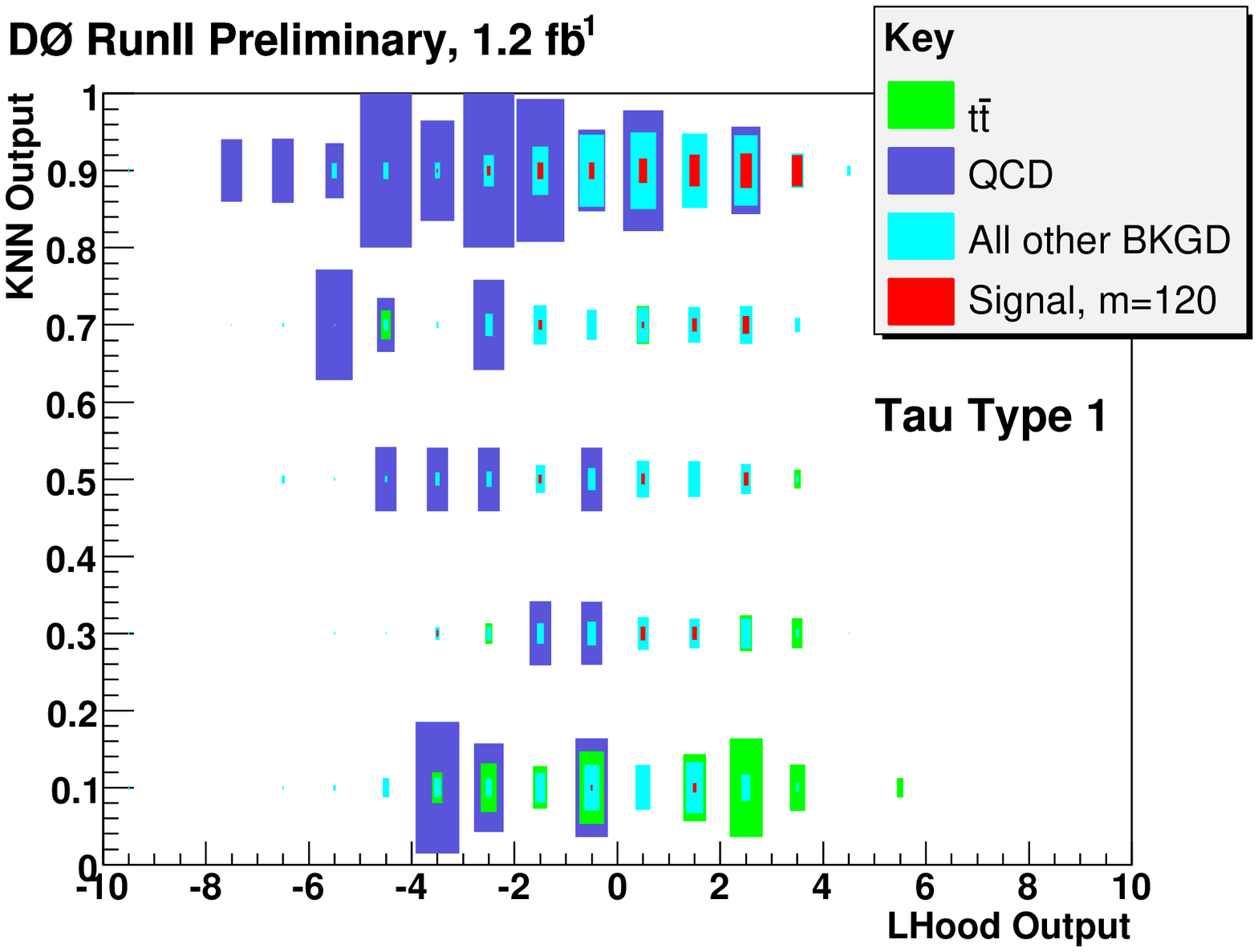}%
\includegraphics[scale=.28]{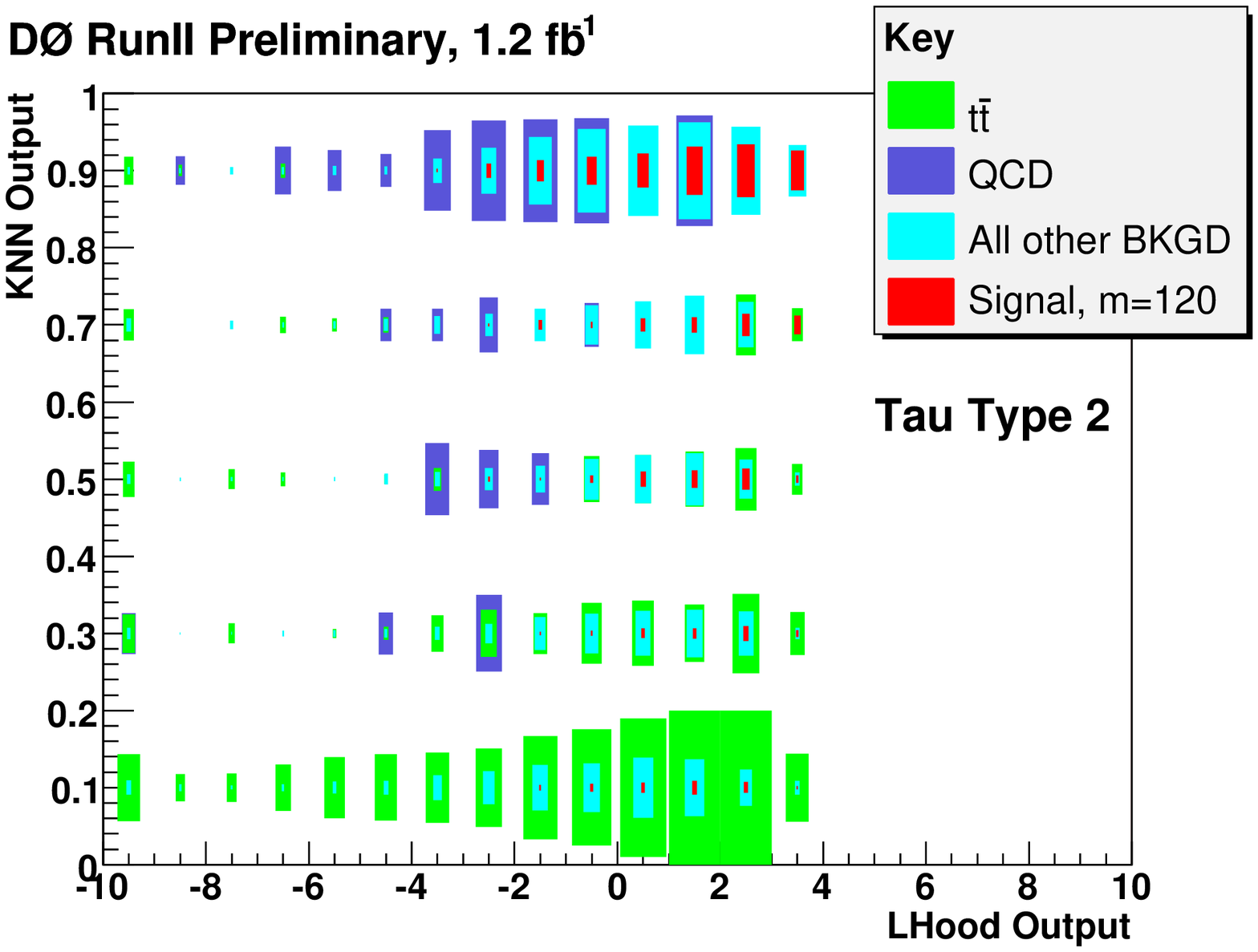}%
\includegraphics[scale=.28]{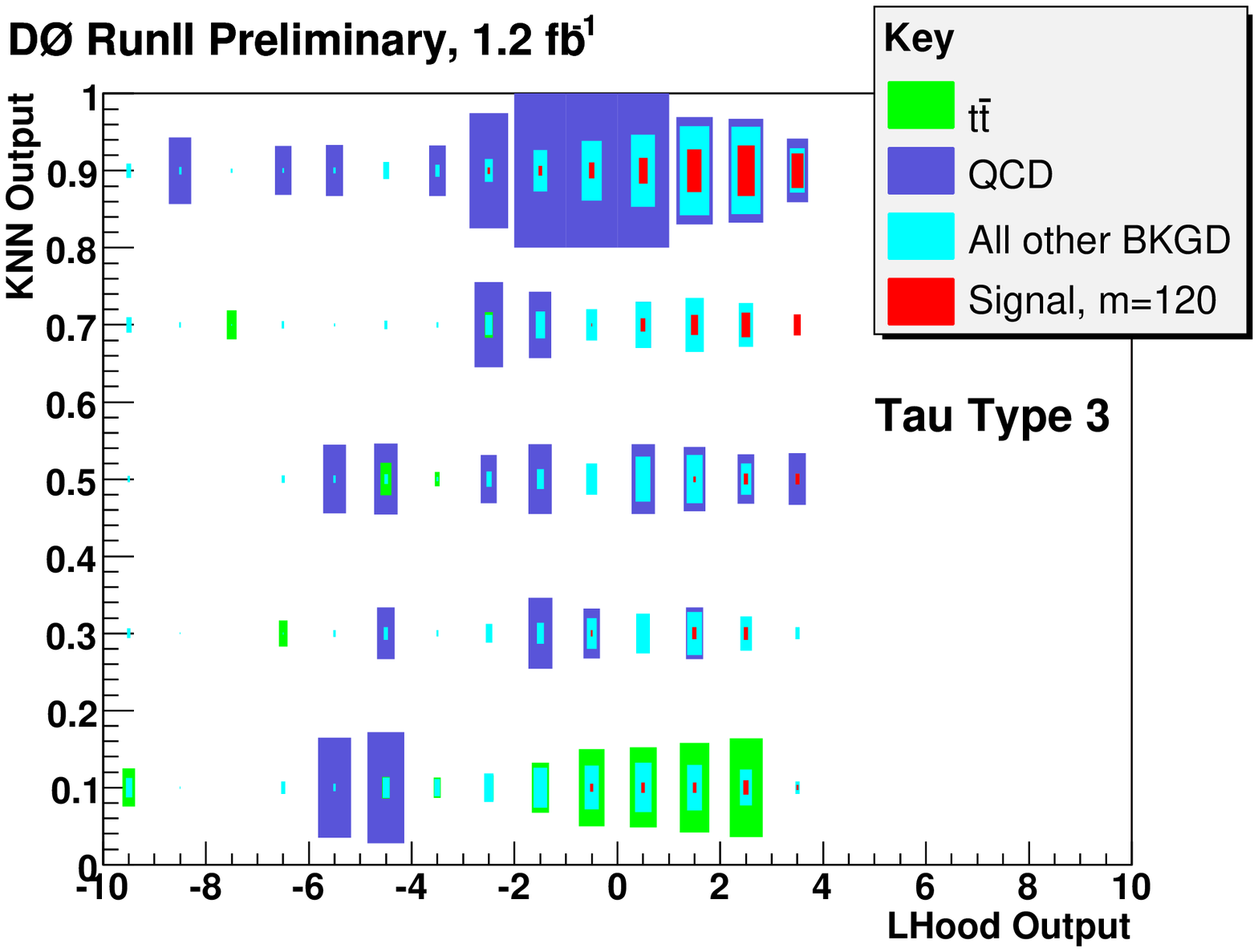}
\caption{KNN output vs. QCD likelihood output with 120 GeV/$c^{2}$ signal after $b$-tagging for: Type 1 taus (left), Type 2 (centre), Type 3 (right.) \label{LHOOS2dbtag}}

\end{figure}

After $b$-tagging our data sample is dominated by \ttbar\ and multijet events.  We employ two multivariate techniques to reject each of the two leading backgrounds.  To reject \ttbar\ background we apply a
 Kinematic Neural Network (KNN) originally developed in \cite{b-hbtau1}.  It
 uses the number of jets in the events, the sum of the transverse momenta of the
 jets (HT), the energy from the four-momentum sum of the muon, tau and jets,
 and the $\Delta \phi$ between the muon and tau candidate as input variables.
 A KNN cut of 0.3 typically offers $\approx 75\%$ rejection in \ttbar\ with
 only a $\approx4\%$ signal loss.  To reject QCD we apply a simple unbinned
 log-likelihood ratio, trained separately for each signal mass point.  We
 consider muon \pT , tau \pT , $\Delta R(\mu,\tau)$, $\mu-\tau$ invariant mass,
 and $(\mu, \tau,\mett)$ invariant mass, or visible mass, as input
 variables, and compute the likelihood of an event to be QCD-like or signal-like in all five variables   Figure \ref{LHOOS2dbtag} shows the KNN and QCD likelihood distributions for signal and background after $b$-tagging.     

\section{Limits and Conclusion}

\begin{figure}[htb]
\centering
\includegraphics[scale=.4]{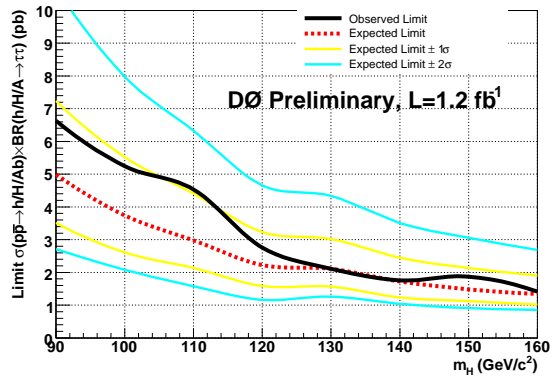}
\caption{ Expected and observed cross section $\times$ BR limits vs. Higgs mass.\label{xseclimits}}
\end{figure}

\begin{figure}[htb]
\centering
\includegraphics[scale=.3]{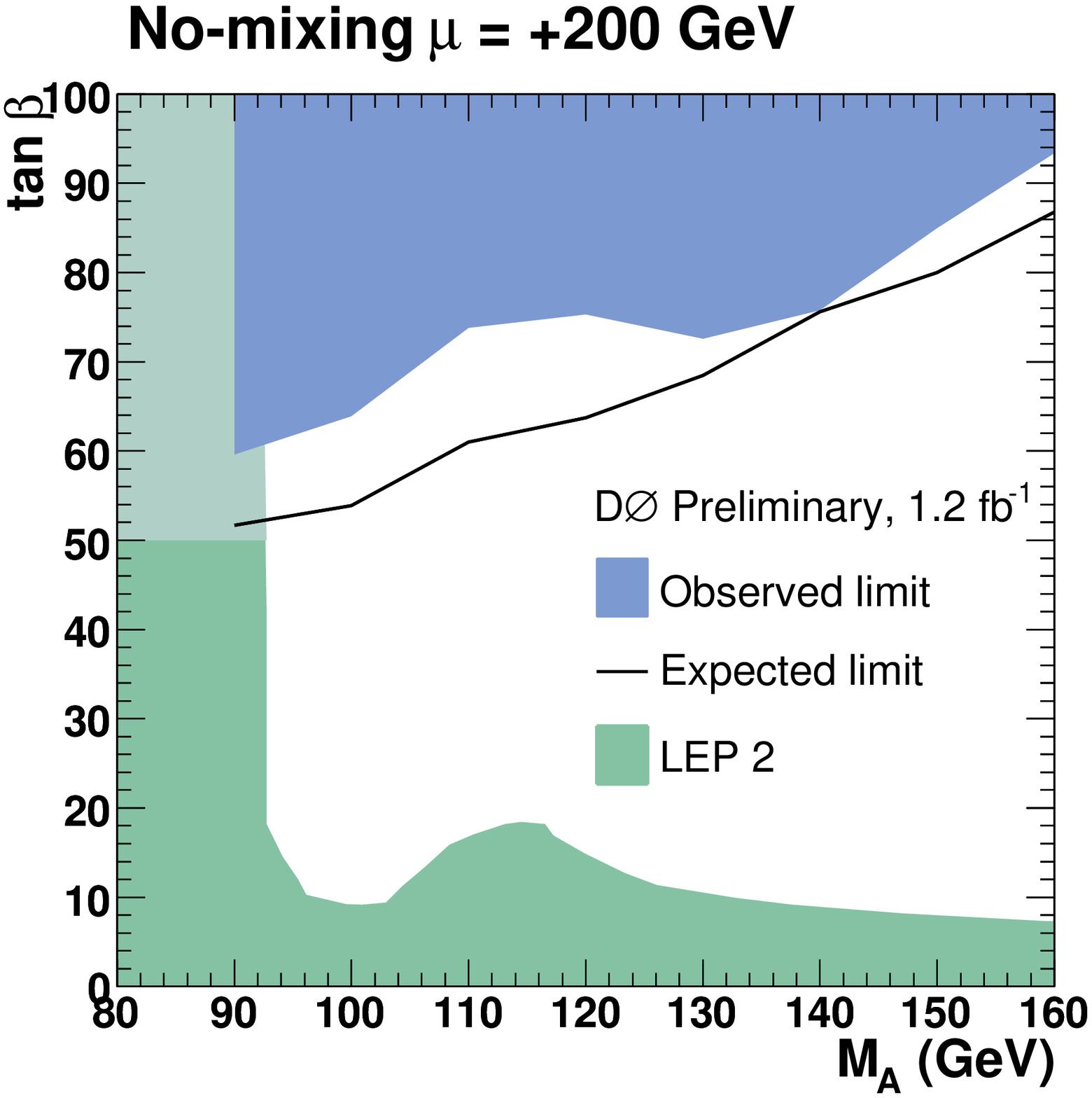}%
\includegraphics[scale=.3]{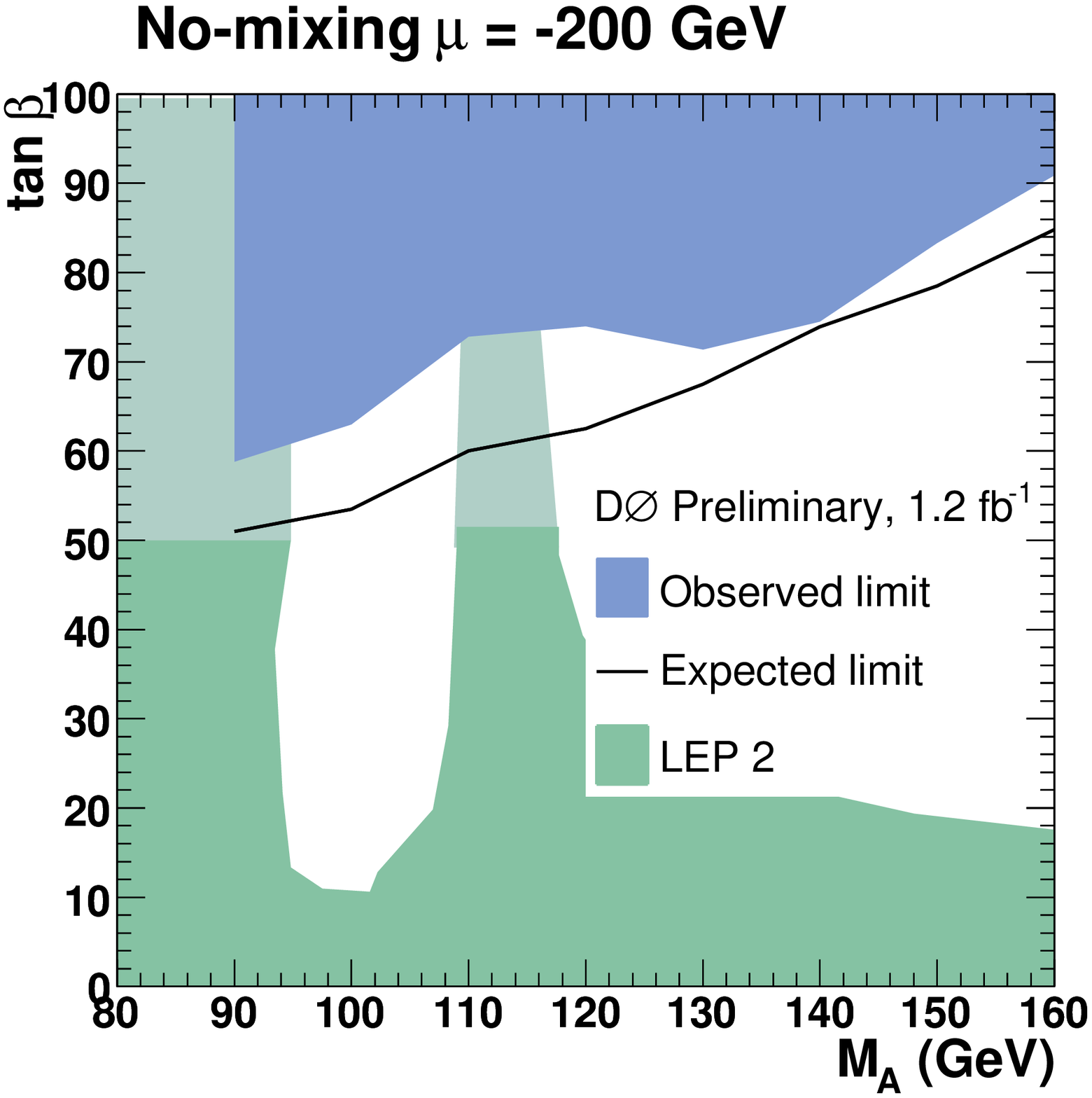}%
\vspace{1mm}
\includegraphics[scale=.3]{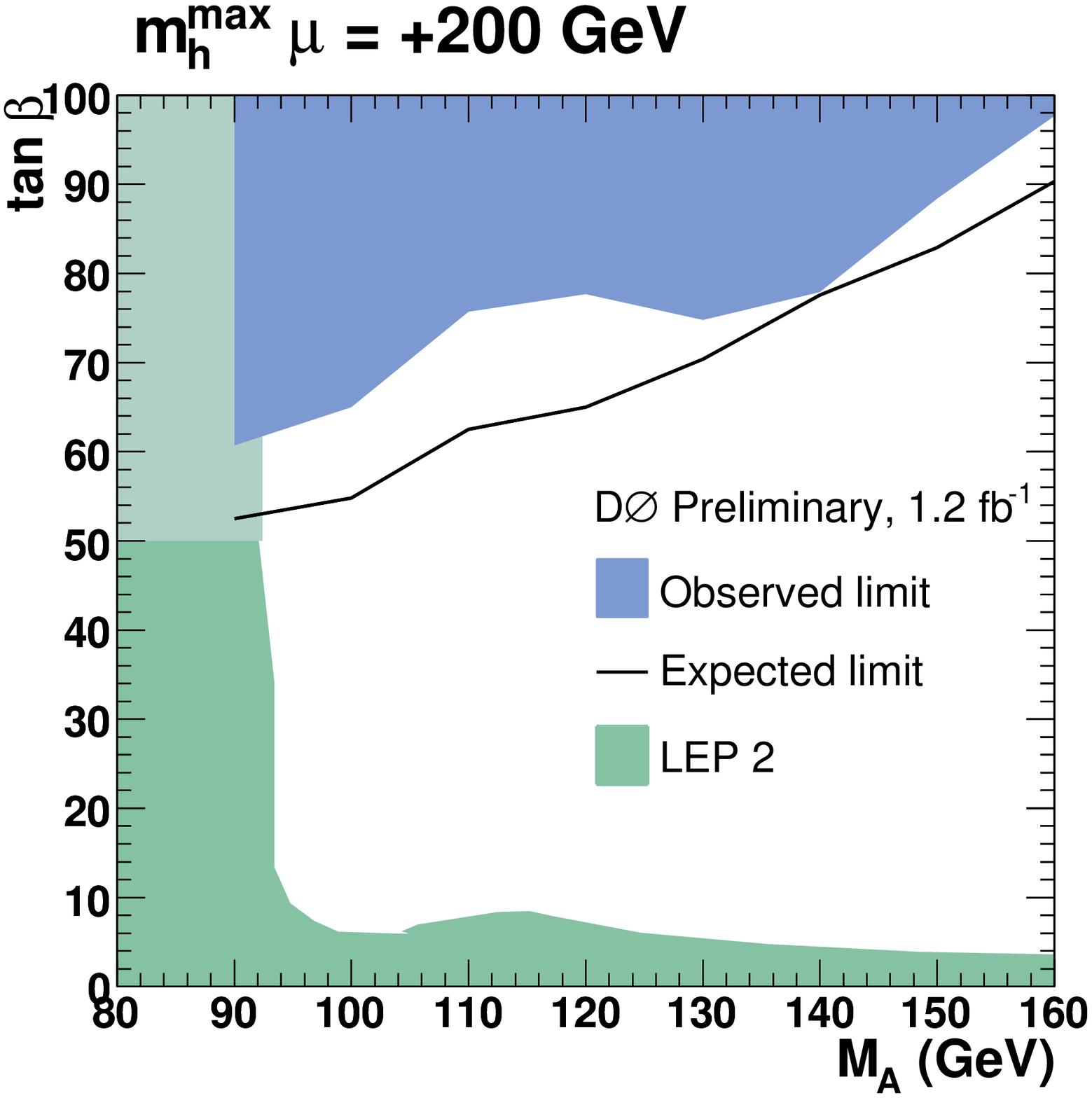}%
\includegraphics[scale=.3]{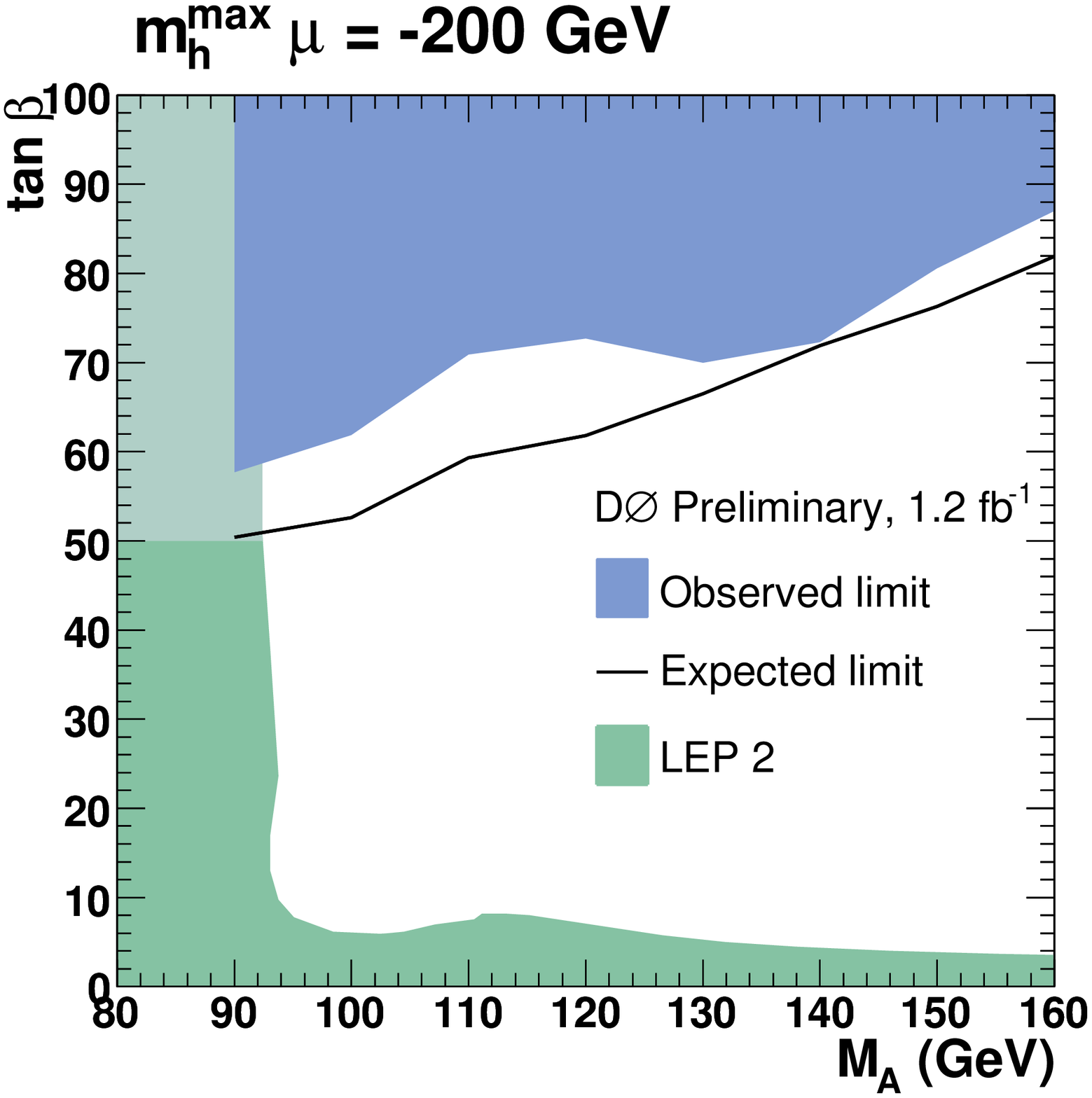}%
\caption{Limits on \tanbeta\ vs. $M_{A}$ for the no-mixing, $\mu > 0$ case (upper left); no-mixing, $\mu < 0$ case (upper right); maximal-mixing, $\mu > 0$ case (lower left); maximal-mixing, $\mu < 0$ case (lower right).}
\label{tanbetalimits}
\end{figure}

We cut on both the KNN and QCD likelihood in the 2D distributions of Fig.~\ref{LHOOS2dbtag}, using expected 
significance as the optimising variable, to determine the cross section limit.  Once we have the cross section limits, shown in figure \ref{xseclimits}, we use FeynHiggs v2.6.2 \cite{b-feynhiggs,b-feynhiggs2,b-feynhiggs3,b-feynhiggs4} 
to interpret the limits in four Minimal Supersymmetric Standard Model (MSSM) scenarios: with no $m_{h}$-mixing and maximal $m_{h}$-mixing, 
and with $\mu = \pm 200$ GeV/$c^{2}$.  Fig.~\ref{tanbetalimits} shows the expected and observed limits in the $\tan \beta$ vs. $m_A$ plane.  
For $m_{A}$ between 90 and 160 GeV/$c^{2}$ we can exclude \tanbeta\ above 100 in all four scenarios at all mass points.

We have performed a search for associated neutral Higgs boson production in the \btautau\ final state with 1.2fb$^{-1}$ of data with the D0 Detector at Fermilab.  In the absence of significant signal we set cross section times branching ratio limits on $\phi + b$ production and interpret the result in several benchmark MSSM scenarios.  The search is complimentary to \bbb\ searches and provides enhanced sensitivity in models where $m_{\phi} \sim M_{Z}$.

\section*{References}
\bibliography{herner}

\end{document}